\documentclass[%
 reprint,
 amsmath,amssymb,
 aps,
]{revtex4-2}
\usepackage{amsmath}
\usepackage{graphicx}
\usepackage{graphics}
\usepackage{braket}
\graphicspath{{Images/}}
\usepackage{graphicx,caption}
\usepackage{subcaption}
\DeclareMathOperator{\Tr}{Tr}

\begin{document}

\title{Floquet-Lindblad Master Equation Approach to Open Quantum System Dynamics}

\author{Fenton Clawson}
\author{Edward B. Flagg}
\affiliation{Department of Physics and Astronomy, West Virginia University, Morgantown, WV, 26506 USA}

\begin{abstract}
    A new solver, named FLiMESolve, is exhibited which is well suited for time-periodic (Floquet-type) Hamiltonians. This solver maintains the benefits of other solvers presently considered, while superceding their usefulness within its own regime. FLiMESolve is able to efficiently and accurately simulate time-periodic systems while being faster and more accurate than other solvers and with the ability to relax the secular approximation inherent to the Lindblad Equation.
\end{abstract}

\maketitle

\section{Introduction}
Computational physics has been a backbone of experimental and theoretical physics since the 1940's, when the first nuclear bomb and ballistic simulations were performed at Los Alamos Laboratory \cite{barberBallisticiansWarPeace1955}. Later on in the same decade, the first Monte Carlo methods were created, becoming one of the most ubiquitous and important simulation methods that is still in use today \cite{haighLostAlamosBets2014}. Later, in 1976, Goran Lindblad created what is now known as the Lindblad equation, opening a whole field of study into dissipative or ``open'' quantum systems \cite{lindbladGeneratorsQuantumDynamical1976a}. Nowadays, simulation of open quantum systems has been formalized into coding packages and user-friendly functions, such as QuTiP or Qiskit, which allow users to simulate these systems without having the in-depth knowledge required to build the computational backbone from the ground-up \cite{johanssonQuTiPOpensourcePython2012b,johanssonQuTiPPythonFramework2013b,javadiabhariQuantumComputingQiskit2024}. QuTiP specifically is described as an ``open-source software package for the simulation of open quantum systems.'' QuTiP provides several solvers with different regimes of validity, such as its MESolve function, which is a good first-pass simulation tool that utilizes the Lindblad equation to simulate open quantum systems. Also within QuTiP is FMMESolve, a solver that is ostensibly meant to be well-suited to the Floquet regime of ODEs, i.e., those with time-periodic coefficients.

Research on open quantum systems in the regime of validity of Floquet theory is often focused on finding solutions to non-Lindblad-form quantum master equations \cite{shirleySolutionSchrodingerEquation1965, blumelDynamicalLocalizationMicrowave1991, grifoniDrivenQuantumTunneling1998}. The disadvantage of these approaches is the necessity of using a spectral density function, which describes the density of states of the environment(s) coupled to the system \cite{leggetQuantumTunnelingPresence1984,leggettDynamicsDissipativeTwostate1987}. The Lindblad master equation, on the other hand, is a useful tool for calculating the time evolution of open quantum systems in that additional environmental couplings can be considered by the creation of additional Lindblad-type terms \cite{brasilSimpleDerivationLindblad2013}. Complexity is reduced in that the spectral density function may be replaced by a single observable decay rate. This is suitable for modeling experimental results, wherein the spectral density function cannot be directly measured but a decay rate can be. The Floquet-Lindblad master equation as used for the simulation of open quantum systems has seen some interest for this reason \cite{ikedaGeneralDescriptionNonequilibrium2020,ikedaNonequilibriumSteadyStates2021}.

Here we describe and demonstrate a new solver for QuTiP named FLiMESolve that is suited for the Floquet regime of ordinary differential equations. This solver avoids some of the pitfalls and complications of MESolve and FMMESolve. First, it is specifically designed for time-periodic Hamiltonians, which leads to a solver that is multiple times faster than MESolve for problems in the Floquet regime. Furthermore, FLiMESolve makes use of the more empirically motivated decay rates, as opposed to spectral density functions, making it more accessible than FMMESolve for beginners. Finally, as shown by simulations of example systems, FLiMESolve can go beyond the secular approximation, allowing it to deal with periodic Hamiltonians with a time dependence more complex than a single frequency while still keeping a high computational efficiency.

\section{Theoretical Considerations}
%
%The derivation will somewhat follow 
% The derivation in this section will parallel those found elsewhere \cite{brasilSimpleDerivationLindblad2013,ikedaGeneralDescriptionNonequilibrium2020,ikedaNonequilibriumSteadyStates2021} and is given both to illustrate the differences in this derivation and to provide a clear picture of the exact theory underpinning the new computational tool presented in this paper. The form of the master equation as presented in this paper is not the usual Floquet-Lindblad Master equation in that the Floquet decomposition of the system operators are kept within the operators themselves, rather than being defined as sums on the ``top level" Lindbladian. This has some computational convenience, and will be discussed as it arises.

\subsection{Lindblad Master Equation with Floquet Operators}
The starting point of this derivation is the time-local Born-Markov master equation, as is common with Lindblad type master equations \cite{brasilSimpleDerivationLindblad2013}. A Hamiltonian is defined, containing system, bath, and system-bath interaction parts, as:
\begin{equation} \label{EQ:Hamiltonian_def}
\hat{H} = \hat{H}_{S}(t)+\hat{H}_B+\hat{H}_I
\end{equation}
For brevity, in the following, $\hbar$ is taken to be one. The Born-Markov master equation describes the time evolution of the density matrix operator of the open system by tracing over the states of the bath:
\begin{equation} \label{EQ:Born-Markov}
\frac{d}{dt} \hat{\rho}_{S} = -\Tr_B\{[\hat{H}_I(t),\int_{t_0}^{t}[\hat{H}_I(t'),\hat{\rho}_S(t)\otimes \hat{\rho}_B]dt']\}.
\end{equation}
In the above, $\hat{H}_I$, $\hat{\rho}_S$, $\hat{\rho}_B$ refer to the interaction picture Hamiltonian and system and bath density matrices in the interaction picture, respectively. Further, the time $t_0$ is some initial time of the system. Additionally, the interaction picture itself is interpreted to be the Floquet basis, i.e. moving to the interaction picture is the same as moving to the Floquet basis. 
% Here is a good place to explain the parallel and connections between the transformation to the interaction picture and the transformation to the Floquet basis. This is a note by Ned to remind us to write that paragraph.

The first step in moving from the Born-Markov master equation to the Floquet-Lindblad master equation is to fully expand the system-bath interaction Hamiltonian, $\hat{H}_{I}$. The decomposed interaction Hamiltonian in the interaction picture $\hat{H}_I(t)$ is given by the tensor product of the system operators, $\hat{S}(t)$, and the bath operators, $\hat{B}(t)$, as
\begin{equation} \label{EQ:Decomp_Int}
\hat{H}_{I}(t) =   S(t) \otimes B(t) 
\end{equation}
%
%The determinant of time-evolution operator can be defined as the operator $U(t,t_0)$ given by
%
%\begin{equation} \label{EQ:Time_Evol_Op}
%\operatorname{det}[\hat{U}(t,t_0)]=\operatorname{det}[\hat{U}_S(t,t_0)] =e^{i\int_{t_0}^{t}\mathrm{Tr}[\hat{H}_S(t)] %dt}
%\end{equation},
%
The unitary time-evolution operator which evolves the system from time $t_0$ to time $t$ and without dissipation, defined as $U(t,t_0)=U_{s}(t,t_0) \otimes U_{b}(t,t_0) = U_{s}(t,t_0) \otimes e^{i\hat{H}_Bt}$ can be used to move the system operators $\hat{S}$ and bath operators $\hat{B}$ to the interaction picture as
\begin{equation} \label{EQ:Int_Def1}
S(t) = U^{\dagger}(t,t_0)S U(t,t_0)
\end{equation}
and
\begin{equation} \label{EQ:Int_Def2}
\hat{B}(t) =  e^{i\hat{H}_Bt}\hat{B}e^{-i\hat{H}_Bt}
\end{equation}
Using this decomposition in the Born-Markov master equation (Eqn.~($\ref{EQ:Born-Markov}$), above) gives:
\begin{equation} \label{EQ:BM-Decomp}
\begin{split}
\frac{d}{dt} \hat{\rho}_{S} = -\Tr_B\{[ S(t) \otimes &B(t), \\
 \int_{t_0}^{t}[S(t') \otimes B(t'),\hat{\rho}_S(t)&\otimes \hat{\rho}_B]dt']\}
\end{split}
\end{equation}

Next, it's assumed that the system Hamiltonian is periodic in time with some fundamental frequency $\omega$. In a real open quantum system, this is usually accomplished through the application of time varying electric or magnetic field, e.g., laser light or microwave fields. Floquet's theorem asserts that the general form of the solution of a linear differential equation, like the Schr\"odinger equation, with coefficients that are periodic takes the form \cite{shirleySolutionSchrodingerEquation1965}
\begin{equation} \label{EQ:Floq-Def}
F(t) = \Phi(t)e^{-iQt}.
\end{equation}
In the above, $\Phi(t)$ is a matrix of periodic functions with the same periodicity as the coefficients in the differential equation, and $Q$ is a constant diagonal matrix. Then, because the Hamiltonian is Hermitian, $F(t)$ can be expressed as a unitary operator and the time-evolution operator $U(t,t_0)$ can be expressed as
\begin{equation} \label{EQ:TE_Op_Def}
U(t,t_{0}) = F(t)F^{-1}(t_{0})
\end{equation}
The above can be re-expressed in terms of the Floquet states $\ket{\Psi_\beta}$ and modes $\ket{\phi_\beta}$
which correspond to the column vectors of the matrix $F(t)$ and $\Phi(t)$, respectively \cite{shirleySolutionSchrodingerEquation1965,grifoniDrivenQuantumTunneling1998}:
\begin{equation} \label{EQ:Floq_States_Def}
\begin{split}
\ket{\Psi_{\beta}(t)} = &\ket{\phi_{\beta}(t)}e^{-i\epsilon_{\beta}t} \\
\textrm{with}\: (\ket{\Psi_0(t)}\ket{\Psi_1(t)}&...\ket{\Psi_{N}(t)})=F(t)
\end{split}
\end{equation}
In Eqn.~(\ref{EQ:Floq_States_Def}), $N$ is the dimensionality of the system, and $\epsilon_\beta$ refers to a Floquet quasienergy of the system. These quasinergies are the entries of the diagonal matrix defined in Eq.~(\ref{EQ:Floq-Def}). Using the above, the unitary time-evolution operator for the system Hamiltonian can be rewritten as
\begin{equation} \label{EQ:Floq_Unitary_Op_Def}
U(t,t_{0}) 
    =\sum_{\beta}\ket{\Psi_{\beta}(t)}\bra{\Psi_{\beta}(0)}
    =e^{-i\epsilon_{\beta}t}\sum_{\beta}\ket{\phi_{\beta}(t)}\bra{\phi_{\beta}(0)}
\end{equation}
This form of the time-evolution operator can be used to re-express the system operators in the Floquet basis as:
\begin{equation} \label{EQ:Gen_Ops_Floq_basis}
\begin{aligned}
\hat{S}(t) =& \sum_{\alpha,\beta}\bra{\Psi_{\alpha}(t)}\hat{S}\ket{\Psi_{\beta}(t)}\ket{\Psi_{\alpha}(0)}\bra{\Psi_{\beta}(0)}\\
=& \sum_{\alpha,\beta}e^{i(\epsilon_{\alpha}-\epsilon_{\beta})t}\bra{\phi_{\alpha}(t)}\hat{S}\ket{\phi_{\beta}(t)}\ket{\phi_{\alpha}(0)}\bra{\phi_{\beta}(0)}
\end{aligned}
\end{equation}
%
%A single index pair ($\alpha,\beta$) in the equation above defines a single matrix element of the operator in the Floquet basis as
%
%\begin{equation} \label{EQ:Sys_Ops_Floq_One_Index}
%\begin{aligned}
%S_{\alpha,\beta}(t) = &\bra{\Psi_{\alpha}(t)} \hat{S} \ket{\Psi_{\beta}(t)}
%\end{aligned}
%\end{equation}
%
%Further, equation \ref{EQ:Gen_Ops_Floq_basis} defines a transition operator that can be expressed in the Floquet state and mode basis, respectively, as
%
%\begin{equation} \label{EQ:Sys_Ops_Floq_One_Index}
%\begin{aligned}
%\kappa_{\alpha,\beta}(t) = \ket{\Psi_{\alpha}(0)}\bra{\Psi_{\beta}(0)}=& e^{i(\epsilon_{\alpha}-%\epsilon_{\beta})t}\ket{\phi_{\alpha}(0)}\bra{\phi_{\beta}(0)}
%\end{aligned}
%\end{equation}

To get to the final form of the operators used in the present work, two more operations are necessary. First, the periodicity of the field-interaction Hamiltonian means that the expectation value in Eqn.~(\ref{EQ:Gen_Ops_Floq_basis}) can be re-expressed using a Fourier series of harmonics of the system frequency
\begin{equation}\label{EQ:Op_Fourier_Expansion}
    \bra{\phi_\alpha(t)} \hat{S} \ket{ \phi_\beta(t)}=\sum_{k=-\infty}^{\infty} e^{i k \omega t} \mathcal{S}_{\alpha,\beta}(k)
\end{equation}
In the above equation, the coefficient $\mathcal{S}_{\alpha,\beta}(k)$ is the $k^{th}$ Fourier coefficient of the $\alpha,\beta$ matrix element of the system operator in the interaction picture. Secondly, operators are explicitly taken to be in the Floquet mode basis. This means that a new operator $\hat{S}_{\alpha,\beta}^{k}(t)$ can be defined as
\begin{equation}\label{EQ:FLoq_Op_Mode_Defs}
    \hat{S}_{\alpha,\beta}^{k}(t)= e^{i\Delta_{\alpha,\beta}^{(k)} \omega t} \mathcal{S}_{\alpha \beta}(k)\ket{\phi_{\alpha}(0)}\bra{\phi_{\beta}(0)}
\end{equation}
In the above, $\Delta_{\alpha,\beta}^{(k)} = \frac{\epsilon_{\alpha}}{\omega}-\frac{\epsilon_{\beta}}{\omega}+ k$. This operator can be understood as the Fourier series form of a matrix element in the Floquet mode basis of the system operator $\hat{S}$. Conveniently, the Floquet representation of these operators allows for a concise matrix expression of these operators, which can be used to find a solvable non-secular Lindblad-form master equation.

Returning now to the Born-Markov master equation, the outer commutator of Eqn.~(\ref{EQ:Born-Markov}) can be expanded to yield:
\begin{equation}\label{EQ:Expanded_Borne_Markov}
\begin{aligned}
    \frac{d }{d t}\hat{\rho}_S =  \int_0^\infty d t' 
    &\left[\hat{S}(t')\hat{\rho}_S(t),\hat{S}^\dag(t)\right] 
    \Tr_B\left\{\hat{B}^\dag(t)\hat{B}(t')\hat{\rho}_B\right\} \\
    +&\left[\hat{S}(t),\hat{\rho}_S(t)\hat{S}^\dag(t')\right]  
    \Tr_B\left\{\hat{B}^\dag(t')\hat{B}(t)\hat{\rho}_B\right\} \\
\end{aligned}
\end{equation}
To recover a nonsecular form of the Lindblad equation, two new operators, $\hat{L}^0(t)$ and $\hat{L}^1(t)$, will be defined as \cite{liaoLindbladRedfieldForms2021}
\begin{equation}\label{EQ:Lindblad_Op_Defs}
\begin{aligned}
    \hat{L}^0(t)=&\int_0^{\infty}dt' \hat{S}(t') \Tr_B\left\{\hat{B}(t)\hat{B}(t')\hat{\rho}_B\right\}\\
    \hat{L}^1(t)=&\hat{L}^{0}(t)+\hat{S}(t)
\end{aligned}
\end{equation}
With these operators, and some time-consuming algebra, the equation of motion (\ref{EQ:Expanded_Borne_Markov}) can once again be written in a Lindblad form. Defining the system Hamiltonian as $\hat{H}_S$, the Lamb shift Hamiltonian as $\hat{H}_{LS}$,the system density matrix as $\hat{\rho}_S$, and the dissipator as $\hat{D}(\hat{\rho}_S)$, the Lindblad equation is given as \cite{liaoLindbladRedfieldForms2021}
\begin{equation}\label{EQ:Liao_Lindblad}
\begin{aligned}
    \frac{d}{dt} \hat{\rho}=-i[\hat{H}_S+\hat{H}_{LS},\hat{\rho}_S]+\hat{D}(\hat{\rho}_S)
\end{aligned}
\end{equation}
In the above, the dissipator and Lamb shift terms are defined as \cite{liaoLindbladRedfieldForms2021}:
\begin{widetext}
\begin{equation}\label{EQ:Liao_Diss_and_Lambshift}
\begin{aligned}
    \hat{D}(\hat{\rho}_S)=&\left[\hat{L}^{1}(t) \hat{\hat{\rho}}_S(t) \hat{L}^{1 \dagger}(t)-\frac{1}{2}\{\hat{L}^{1 \dagger}(t) \hat{L}^{1}(t), \hat{\rho}_S(t)\}\right] \\
    -&\left[\hat{L}^{0}(t) \hat{\rho}_S(t) \hat{L}^{0^{\dagger}}(t)-\frac{1}{2}\{\hat{L}^{0 \dagger}(t) \hat{L}^{0}(t), \hat{\rho}_S(t)\}\right] \\
    -&\left[\hat{S}(t) \hat{\rho}_S(t) \hat{S}^{\dagger}(t)-\frac{1}{2}\{\hat{S}^{\dagger}(t) \hat{S}(t), \hat{\rho}_S(t)\}\right] \\
    \hat{H}_{LS} = &\frac{1}{2}[\hat{L}^{0 {\dagger}}(t)\hat{S}(t)-\hat{S}^{\dagger}(t)\hat{L}^{0}(t)]
\end{aligned}
\end{equation}
\end{widetext}
Using the previously defined Floquet decomposition on the system operators, the operators in Eqn.~(\ref{EQ:Lindblad_Op_Defs}) can be re-expressed as
\begin{equation}\label{EQ:Lindblad_Ops_Floquet_Redef}
\begin{aligned}
    &\hat{L}^{0,k}_{\alpha,\beta}(t)= \sum_{\alpha \beta} \Gamma_{\alpha,\beta}^{k}(t) \hat{S}_{\alpha,\beta}^{k}(t), \quad \\
    &\hat{L}^{1,k}_{\alpha,\beta}(t)= \sum_{\alpha \beta}\left(\Gamma_{\alpha,\beta}^{k}(t)+1\right) \hat{S}_{\alpha,\beta}^{k}(t),  \\
    \text{with}&\; \Gamma_{\alpha,\beta}^{k}(t)=\int_0^{\infty} d s e^{i \Delta_{\alpha,\beta}^{(k)}\omega s} \Tr_B\left\{\hat{B}(t) \hat{B}\left(s\right) \hat{\rho}_B\right\}
\end{aligned}
\end{equation}
Finally, a Markov approximation is made for the bath autocorrelation functions. The autocorrelation function in the last line of Eqn.~\ref{EQ:Lindblad_Ops_Floquet_Redef} can be understood as a Fourier transform. By assuming that the bath correlation times are effectively instant (i.e. making the assumption that the spectral density function of the bath takes the form of white noise), the decay rate of the Lindblad terms in the master equation becomes constant:
\begin{equation}\label{EQ:Constant_Coupling_strength}
    \Gamma_{\alpha,\beta}^{k}(t)=\frac{\gamma}{2} \ \forall \ \alpha,\beta,k,t
\end{equation}
The dissipator term can be written in a Lindblad form:
\begin{equation}\label{EQ:Final_Dissipator_Form}
\begin{gathered}
    \hat{D}(\hat{\rho}_S)=\sum_{\alpha, \alpha^{\prime}} \sum_{\beta, \beta^{\prime}} \gamma[\hat{S}_{\alpha,\beta}^{k}(t) \hat{\rho}(t) \hat{S}_{\alpha',\beta'}^{k'^{\dagger}}(t))\\
    -\frac{1}{2} \hat{S}_{\alpha',\beta'}^{k'^{\dagger}}(t) \hat{S}_{\alpha,\beta}^{k}(t) \hat{\rho}(t)-\frac{1}{2} \hat{\rho}(t) \hat{S}_{\alpha',\beta'}^{k'^{\dagger}}(t) \hat{S}_{\alpha,\beta}^{k}(t)]
\end{gathered}
\end{equation}
\subsection{{The Floquet-Lindblad Rate Matrix, R(t)}}
The final step in the process of finding a computationally-efficient form of the Floquet-Lindblad master equation is to find an expression for the individual matrix elements of Eqn.~($\ref{EQ:Final_Dissipator_Form}$) and rearranging them into a rate matrix form. In order to obtain the first-order ODE form of the master equation, it is necessary to ``unfold'' the density matrices into one-dimensional supervectors. A convenient basis to use for this operation is the Floquet mode basis at time $t=0$, as discussed previously and which led to Eqn.~(\ref{EQ:FLoq_Op_Mode_Defs}). 
Making this transformation, and after another round of time-consuming algebra, the the form of the master equation given by $\frac{d}{dt}\hat{\rho}_S(t)=\hat{R}(t)\hat{\rho}_S(t)$ is obtained, with the matrix $\hat{R}(t)$ having elements given by:
\begin{widetext}
\begin{equation}\label{EQ:TDFLiMEFinal}
\begin{aligned}
R{ }_{m n, p q}(t)  =\sum_{\alpha \beta \alpha^{\prime} \beta^{\prime}} \sum_{k k^{\prime}=-\infty}^{\infty} & S_{\alpha \beta}(k) S_{\alpha^{\prime} \beta^{\prime}}^{*}\left(k^{\prime}\right) e^{i\left(\Delta_{\alpha \beta}^{(k)}-\Delta_{\alpha^{\prime} \beta^{\prime}} ^{(k')}\right) \omega t} \\
& \times \left[\delta_{n \alpha} \delta_{\alpha^{\prime} m} \delta_{p \beta} \delta_{q \beta^{\prime}}-\frac{1}{2}\left(\delta_{n \beta^{\prime}} \delta_{\alpha^{\prime} \alpha} \delta_{p \beta} \delta_{q m}+\delta_{m \beta^{\prime}} \delta_{\alpha^{\prime} \alpha} \delta_{p n} \delta_{q \beta^{\prime}}\right)\right]
\end{aligned}
\end{equation}
\end{widetext}
Note that in equation (\ref{EQ:TDFLiMEFinal}), the rate matrix $R_{mn,pq}(t)$ is four-indexed, but only two dimensional.  Given that the density matrix $\hat{\rho}$ is indexed by $n,m$, and that the dimensionality of the system is (again) given by $N$, the unfolding process is given as
\begin{equation}\label{EQ:Mat_vec_trans}
\begin{gathered}
    \hat{\rho}_{n,m} = \Vec{\rho}_{n+Nm}
\end{gathered}
\end{equation}
In Eqn.~(\ref{EQ:Mat_vec_trans}), $\hat{\rho}$ explicitly refers to the matrix form of the density matrix, and $\vec{\rho}$ refers to the supervector form of the density matrix. Using this notation, the first-order ODE given by equation (\ref{EQ:TDFLiMEFinal}) can be written with Einstein notation as
\begin{equation}\label{EQ:FLiME_ODE_form}
\begin{gathered}
    \dot{\vec{\rho}}_{(n+B m)} = \hat{R}_{(m+B n),(p+B q)}(t)\vec{\rho}_{(p+B q)}
\end{gathered}
\end{equation}
The above form of the Master Equation is then solvable as a first-order ODE using QuTiP's ODE infrastructure to evolve suitable systems forward in time in an efficient manner. This efficiency lends itself well to systems with long-time dissipation dynamics but short-time system dynamics. Additionally, the time-dependent nature of Eqn.~(\ref{EQ:FLiME_ODE_form}) allows for the relaxation of the secular approximation. Several examples are shown in the next section to illustrate these advantages.

\subsection{{Secular Approximation and the Negligibility Factor}}
Eqn.~\ref{EQ:TDFLiMEFinal} contains within it all time-dependent \textit{and} independent dynamics of any given system. However, calculating the solution to the \textbf{full} time-dependent Floquet-Lindblad equation can be extremely time consuming and is unnecessary for many applications. For the sake of efficiency, the so-called "negligibility factor" is used. This negligibility factor is defined as:
\begin{equation}\label{NegFac}
    \mathbb{N}_{\alpha,\alpha',\beta,\beta'}^{k,k'} = \frac{\left(\Delta_{\alpha \beta}^k-\Delta_{\alpha^{\prime} \beta^{\prime}}^{k'}\right) \omega}{ \mathcal{S}_{\alpha \beta}^{(k)} \mathcal{S}_{\alpha' \beta'}^{(k')}}
\end{equation}
The numerator of Eq.~(\ref{NegFac}) is the frequency difference found in the exponential term of Eq.~(\ref{EQ:TDFLiMEFinal}). This term, with $\Delta_{\alpha \beta}^k=\left(\frac{\epsilon_{\alpha}}{\omega}-\frac{\epsilon_{\beta}}{\omega}+k\right)$ as the difference in rotation frequency of the two three-indexed system operator terms. This difference in frequencies is itself a frequency, which defines how quickly a term in Eq.~(\ref{EQ:TDFLiMEFinal}) rotates through the complex plane. If a term rotates too quickly, then the effect of the rate product is effectively averaged out over any significant period of time.

The denominator of Eq.~(\ref{EQ:TDFLiMEFinal}), which is the product of two three-indexed system operators decomposed along a single frequency (with the harmonic multiple denoted by $k$), and two state indices ($\alpha$ and $\beta$ denote the Floquet modes by which the operator is sandwiched), describes the magnitude of the effect of the full Fourier-expanded collapse operator product (which is indexed by $\alpha,\alpha',\beta,\beta',k,k'$) on the system. A larger magnitude has a larger affect on the system, and a smaller magnitude the converse. 

The effect of the quotient is to determine which terms of the overall Floquet-Lindblad master equation are "important" enough to be kept. A low value for the negligibility factor means that either the frequency is low or the rate product is large, corresponding to a term that appreciably affects the Hamiltonian on appreciable timescales. Conversely, a large value in the negligibility factor means that either the term rotates very quickly or that the rate product is small, leading to a term that will average out on appreciable timescales. An example of where the negligibility factor becomes important can be seen in the final example system below.

\section{Example Systems}
FLiMESolve is a particularly useful computational tool for systems that fall into the Floquet regime of validity, i.e., those that are driven cyclically in time. In this section, two particular aspects of the utility of FLiMESolve are explored by simulating two systems. The first is the driven 2-level system (2LS) without the rotating wave approximation, for use in exploring the steady state solver. The second is a 2LS driven by repeating $\pi$-pulses, for use in exploring the relaxed secular approximation.

\subsection{Driven Two-Level System}
The first exemplar of the utility of FLiMESolve can be found in a relatively simple system: the driven 2-level system. Specifically considered here is a scenario wherein an oscillating electric or magnetic field is driving a general 2LS, such as an atom, a quantum dot, or a superconducting  qubit. The Hamiltonian for this system can be written using quantum optics notation (and with $\hbar = 1$) as \cite{ermannJaynesCummingsModelMonochromatic2020, caoQubitStronglyCoupled2011, divincenzoRigorousBornApproximation2005, irishGeneralizedRotatingwaveApproximation2007, saikoEmissionSpectrumQubit2018,scullyQuantumOptics1997}:
\begin{equation} \label{EQ:D2LS_Hamiltonian}
\hat{H}(t) = \frac{\omega_0}{2}\hat{\sigma}_{z} + \Vec{E}(t) \cdot{} \hat{\vec{d}}.
\end{equation}
In the above, $\omega_0$ is the transition frequency of the system, $\hat{\sigma}_{z}$ is the Pauli z operator, $\Vec{E}(t)$, is the electric field of the driving laser, and $\hat{\vec{d}}$ is the transition dipole moment of the 2LS. In the following, we will assume that the electric field has the form
\begin{equation} \label{EQ:E-field}
    \vec{E}(t) = \vec{E} \cos (\omega t + \phi)
               = \left( \vec{\epsilon} \mathcal{E} e^{i \omega t}
                       +\vec{\epsilon}^* \mathcal{E}^* e^{-i \omega t} \right)
\end{equation}
where $\vec{\epsilon}$ is the (in general complex) polarization vector, and $\mathcal{E}$ is the complex electric field amplitude.

At low driving powers, for the sake of computational or analytical tractability this system is usually handled well by taking the rotating wave approximation (RWA) \cite{shirleySolutionSchrodingerEquation1965,blochMagneticResonanceNonrotating1940,koteswaraNonlinearDynamicsTwolevel2017,ermannJaynesCummingsModelMonochromatic2020}. Under the RWA, the terms that oscillate very quickly are assumed to have little effect on the system dynamics. The RWA Hamiltonian is given in the Schrodinger picture as \cite{scullyQuantumOptics1997}:
\begin{equation} \label{EQ:D2LS_RWA}
\hat{H}(t) =\frac{1}{2}\begin{bmatrix}
-\omega_{0} & 0 \\
0       & \omega_{0} 
\end{bmatrix}
+
\frac{1}{2}\begin{bmatrix}
0 & \Omega \\
0 & 0 
\end{bmatrix}e^{i{\omega}t}
+
\frac{1}{2}\begin{bmatrix}
0 & 0 \\
\Omega^* & 0 
\end{bmatrix}e^{-i{\omega}t}
\end{equation}
The Rabi frequency, $\Omega = \mathcal{E} \vec{\epsilon}\cdot \vec{d}$, is the magnitude of the coupling between the forward-rotating laser frequency and the dipole moment. This transformation has the benefit of allowing one to rewrite the Hamiltonian in a time-independent and analytically solvable format in a rotating reference frame. Further, this formulation has a definite steady state in the computational basis, which is given by \cite{mullerResonanceFluorescenceCoherently2007, downingResonanceFluorescenceTwo2023}:
\begin{equation}\label{EQ:RWA_steady statepop}
    n_{t=\infty} = \frac{\Omega^2}{4\Delta^2+\gamma^2+2\Omega^2} 
\end{equation}
In the above equation, $\Delta = \omega - \omega_0 $ is the laser detuning from resonance and $\gamma$ is the spontaneous decay rate of the 2LS.

However, while it is a very good approximation at optical frequencies where the laser coupling strength is usually much less than the resonance frequency, the RWA cannot be applied in systems where the driving strength of the field is an appreciable fraction of the resonance frequency of the system; i.e., where $\Omega\ll\omega$ is not true. When this is the case, the RWA is not valid, as the rapidly oscillating terms contribute substantially to the dynamics. The steady state predicted by the RWA becomes inaccurate, as the system instead approaches a nonequilibrium steady state, i.e., an oscillating state to which all initial states are attracted over time \cite{ikedaGeneralDescriptionNonequilibrium2020,ikedaNonequilibriumSteadyStates2021}. 

To recover accurate system dynamics at large Rabi frequencies, the full (non-RWA) Hamiltonian must be used:
\begin{equation} \label{EQ:D2LS_No_RWA}
\hat{H}(t) =\frac{1}{2}\begin{bmatrix}
-\omega_0 & 0 \\
0 & \omega_0 
\end{bmatrix}
+
\frac{1}{2}\begin{bmatrix}
0 & \Omega \\
\tilde{\Omega}^* & 0 
\end{bmatrix}e^{i{\omega}t}
+
\frac{1}{2}\begin{bmatrix}
0 & \tilde{\Omega} \\
\Omega^* & 0 
\end{bmatrix}e^{-i{\omega}t}.
\end{equation}
The new terms given by $\tilde{\Omega} =  \mathcal{E}^* \vec{\epsilon}^*\cdot \vec{d}$ are the coupling between the dipole moment and the ``counter-rotating'' components of the field. However, when the full system Hamiltonian is used, the equation of motion can no longer be analytically solved because of the time dependence of the Hamiltonian. The time dependence is periodic, however, which is handled efficiently by Floquet theory and FLiMESolve. 

\begin{figure}
    \centering
    \setkeys{Gin}{width=8.0cm, height=10cm,keepaspectratio}
\begin{subfigure}{8cm}
    {\includegraphics{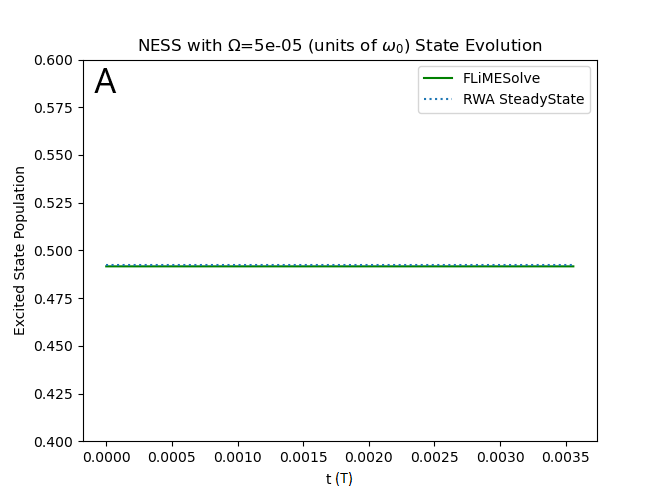 }}
\end{subfigure}
\begin{subfigure}{8cm}
    {\includegraphics{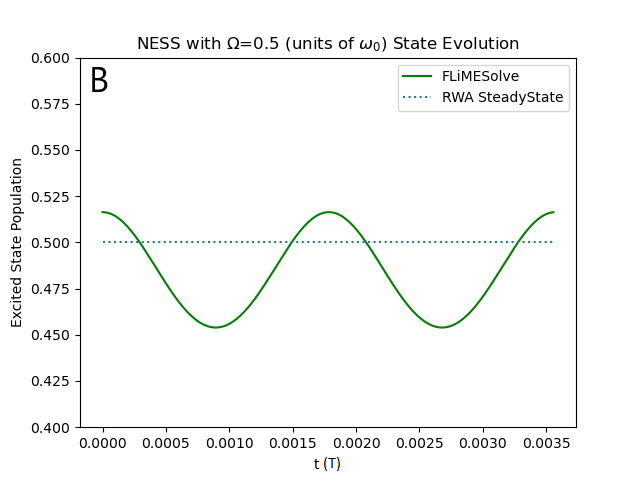}}
\end{subfigure}
        \caption{Time evolution of a driven 2LS with $\Delta=0$, and $\gamma=2.5$ GHz. (a) With a low laser-driving strength ($\Omega/\omega_0=5\times10^{-5}$), a steady state is reached, which agrees well with the RWA steady-state population predicted by Eqn.~(\ref{EQ:RWA_steady statepop}). (b) With the Rabi frequency equal to half the resonance frequency of the 2LS ($\Omega/\omega_0=0.5$), the RWA steady state is no longer a good solution, and the full system dynamics must be accounted for.}
        \label{FIG: LoHiRabiFreq}
\end{figure}

\begin{figure}
  \centering
    \setkeys{Gin}{width=8.0cm, height=10cm,keepaspectratio}
\begin{subfigure}{8cm}
    {\includegraphics{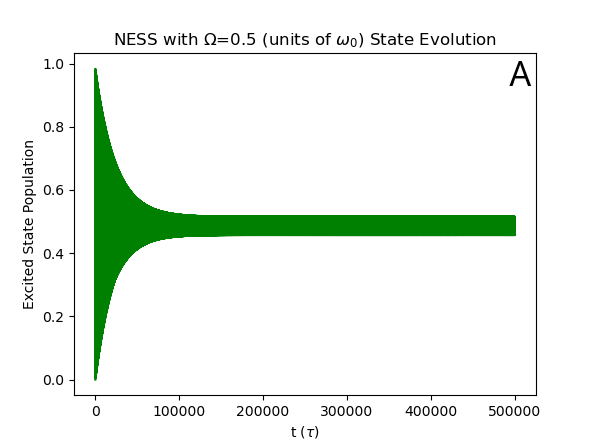}}
\end{subfigure}
\begin{subfigure}{8cm}
    \centering
    {\includegraphics{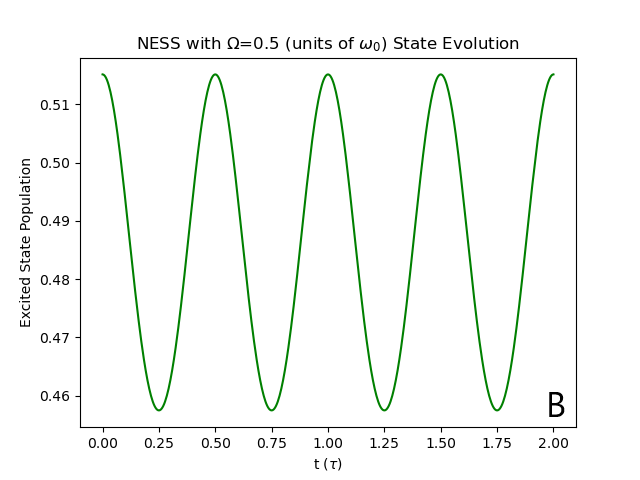}}
\end{subfigure}
        \caption{Plots of the excited state evolution of the resonantly driven 2LS with $\Omega/\omega_0 = 0.5$. (a) The system simulated for 500,000 periods of driving, with the nonequilibrium steady state being reached at around 200,000 periods. (b) Two periods of oscillation in the nonequilibrium steady state, showing the single-period-scale dynamics of the system.}
        \label{FIG: Driven2LS_Excited_State_evol}
\end{figure}

Figure \ref{FIG: LoHiRabiFreq} shows the steady-state time dependence of the excited state of the 2LS under two distinct regimes of driving power. Figure \ref{FIG: LoHiRabiFreq}(a) shows the excited state evolution for very weak driving power ($\Omega/\omega_0=5\times10^{-5}$), and Fig.~\ref{FIG: LoHiRabiFreq}(b) shows the same system under much stronger driving power ($\Omega/\omega_0=0.5$). For the lower driving power in Fig.~\ref{FIG: LoHiRabiFreq}(a), the system evolution calculated by FLiMESolve agrees with the RWA-predicted steady state to less than 1\%, having a very similar population and keeping a constant steady state population throughout the period of driving. Figure \ref{FIG: LoHiRabiFreq}(b), however, shows that under strong driving power the RWA-predicted steady state is no longer accurate. This is due both to the average excited state population deviating from the RWA-predicted steady state, and because the population oscillates throughout the period of driving, which is much different behavior from the constant steady state predicted by the RWA.
Further, because the system often relaxes on a timescale on the order of $10^{5} - 10^{6}$ times longer than the period of the laser, the system must be simulated for a rather large number of periods to reach its nonequilibrium steady state (NESS). Figure \ref{FIG: Driven2LS_Excited_State_evol} shows the evolution of the simulated 2LS when the Rabi frequency is equal to half the resonance frequency of the 2LS. Figure \ref{FIG: Driven2LS_Excited_State_evol}(a) shows the overall decay of the driven 2LS to a nonequilibrium steady state (NESS) over a few hundreds of thousands of periods of driving, while Fig.~\ref{FIG: Driven2LS_Excited_State_evol}(b) shows the single-period dynamics that persist throughout the system evolution and into the NESS.

Since the system must be simulated for so long to reach the nonequilibirum steady state when straightforwardly solving the associated master equation, this results in a very large computational cost. However, as it is uniquely suited to simulating time-periodic systems, FLiMESolve can perform the simulation faster than, e.g., QuTiP's MESolve. This is after some initial overhead investment to solve for the Floquet modes and quasienergies, set up the internal rate matrices in FLiMESolve, perform basis changes to the Floquet basis, etc. Table \ref{TABLE: Table1} shows the solution time, in seconds, required for MESolve versus FLiMESolve for various simulation times. All simulations were performed using using an Intel i7-12700K processor and 16 GB RAM. When the number of periods simulated is low, MESolve is slightly faster than FLiMESolve. As previously mentioned, this is due to the overhead required to set up FLiMESolve. However, as the number of periods simulated grows, FLiMESolve catches up to MESolve in computational time (at around 100 periods of simulated time), before surpassing MESolve in speed. FLiMESolve's advantage appears to saturate at around 5.5-6.0 times faster than MESolve for large simulation times. This means that for time-periodic systems, FLiMESolve is either comparable to or faster than MESolve for all simulation times. However, the exporting of these solutions requires applying QuTiP's ``quantum object'' wrapper, \texttt{QObj()}, around the state individually at every time step. This process reduces the overall speed of FLiMESolve, causing the total run time to be only slightly shorter for FLiMESolve than MESolve. However, FLiMESolve is currently written using numpy for most of its data operations, and a rewriting of FLiMESolve in QuTiP's data layer (which is a faster way to do data operations and which interacts better with the quantum object wrapper) should alleviate this bottleneck and restore the efficiency of FLiMEsolve over MESolve as seen in Table \ref{TABLE: Table1}.

\begin{center}
\begin{table}[t]
\resizebox{\columnwidth}{!}{
\begin{tabular}{||c c c c c||} 
    \hline
    \multicolumn{1}{|p{2cm}|}{\centering Number of Periods} & \multicolumn{1}{|p{2cm}|}{\centering FLiMESolve Solution Time (s)} & \multicolumn{1}{|p{2cm}|}{\centering MESolve  Solution Time (s)} & \multicolumn{1}{|p{2cm}|}{\centering Solution Time Quotient} & \multicolumn{1}{|p{2cm}|}{\centering  Total Run Time Quotient}  \\
    \hline\hline
    $10^{1}$ & $5.0\pm5.3\times10^{-3}$ & $2.0\pm0.3\times10^{-3}$ & 0.4  & 0.39 \\ 
    \hline
    $10^{2}$ & $2.1\pm0.2\times10^{-2}$ &  $1.91\pm3.3\times10^{-1}$ & 8.9  & 2.57 \\ 
    \hline
    $10^{3}$ & $0.04\pm0.002$ & $0.19\pm0.008$ & 4.61 & 1.14  \\ 
    \hline
    $10^{4}$ & $0.30\pm0.012$ & $1.71\pm0.048$ & 5.43 & 1.12 \\
    \hline
    $10^{5}$ & $2.63\pm0.03$ & $17.48\pm0.48\pm$ &  5.92 & 1.18\\
    \hline
    $10^{6}$ & $28.87\pm0.35$ & $181.87\pm2.27$ &  5.62 & 1.23 \\ [1ex] 
    \hline
\end{tabular}
}
 \caption{Solution time and total run time in seconds for FLiMESolve and MESolve for the driven two-level system and varied numbers of periods of simulation. The rightmost two columns show the solution time of MESolve divided by the solution time of FLiMESolve (i.e., before applying the QObj() wrapper to the Floquet states and solution states) and the total run time of MESolve divided by the total run time of FLiMESolve (\textit{after} applying the QObj wrapper to the solution and Floquet states), respectively.}
    \label{TABLE: Table1}
\end{table}
\end{center}
\subsection{Bichromatic Two-level System}

The bichromatic 2LS is an interesting and dynamic system, and motivates the use of FLiMESolve due to the fact that even under the rotating wave approximation, the system cannot be made fully time-independent. As with the excited two level system, we first start with the Atomic Hamiltonian. Assuming that the normalized energy eigenstates of the system are such that $\hat{H}_{atom}\ket{\Psi_1}=E_1\ket{\Psi_1}$ and $\hat{H}_{atom}\ket{\Psi_2}=E_2\ket{\Psi_2}$, the atomic Hamiltonian can be written in the rotating wave approximation as
\begin{equation}\label{EQ:Bichrom_atomic_Ham}
    \hat{H}_{atom} =
    \begin{bmatrix}
    -\Delta & 0 \\
    0 & \Delta
    \end{bmatrix}
\end{equation},
where $\Delta=E_2-E_1$ is the energy difference between the two levels of the system.  Additionally, the same dipole operator is defined as before, namely that
\begin{equation}\label{EQ:Bichrom_Dipole_moment}
    \hat{\vec{d}} = \Vec{d}\ket{0}\bra{1}+\Vec{d}^{*}\ket{1}\bra{0} = \
    \begin{bmatrix}
    0 & \Vec{d} \\
    \Vec{d}^* & 0 
    \end{bmatrix}
\end{equation}
Now it's time to describe the atom-laser interaction. As with before, this is defined as the dot product between the dipole moment of the atom and the electric fields of the laser(s). However, unlike before, there are now two lasers for which to account:
\begin{equation}\label{EQ:Bichrom_Dipole_dot_E}
    \hat{\vec{d}}\bullet\vec{E}(t) = \hat{\vec{d}}\bullet\vec{E}_1(t)+\hat{\vec{d}}\bullet\vec{E}_2(t)
\end{equation}
Now, the act of performing the rotating wave approximation amounts to ignoring the quickly rotating terms in Eqn.~(\ref{EQ:Bichrom_Dipole_dot_E}). In this case, that means ignoring terms with sums in the frequencies, i.e., only keeping terms that depend on the beat frequency. Performing the RWA and then separating the Hamiltonian based on time dependence, the overall Hamiltonian can be written as:
\begin{equation}\label{EQ:Hamiltonian_RotatingFrame_3}
    \begin{aligned}
       \tilde{H}_{0} &= \frac{\Bar{\Delta}}{2} \
        \begin{bmatrix}
            -1 & 0 \\
            0 & 1 
        \end{bmatrix} \\
        \tilde{H}^{+}(t) &=  -\frac{1}{2} \
        \begin{bmatrix}
            0 & \Omega_2 \\
            \Omega_1^* & 0 
        \end{bmatrix} e^{\frac{ib}{2}t} \\
        \tilde{H}^{-}(t) &=  -\frac{1}{2} \
        \begin{bmatrix}
            0 & \Omega_1 \\
            \Omega_2^* & 0 
        \end{bmatrix} e^{-\frac{ib}{2}t}
    \end{aligned}
\end{equation}

The discussion portion of the Bichromatic two level system will solely concern itself with the emission spectra of the system. This is to more clearly show the parallels between simulation results produced by FLiMESolve and the results produced by Chris Gustin et. al. \cite{gustinHighresolutionSpectroscopyQuantum2021}. The simulated system in this simulation will use the same numbers as Gustin's experimental system, i.e., that the resonance of the two level system is at $\frac{\omega_0}{2\pi} = 330 $ THz, and that the lifetime of the system is $455$ ps.

\begin{figure*}
    \centering
    \setkeys{Gin}{width=8.0cm, height=10cm,keepaspectratio}
\begin{subfigure}{8cm}
    {\includegraphics{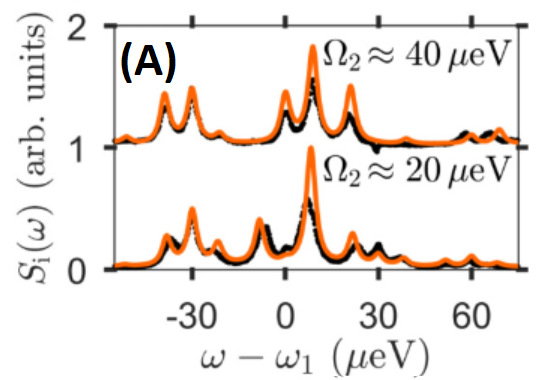}}
\end{subfigure}
\begin{subfigure}{8cm}
    {\includegraphics{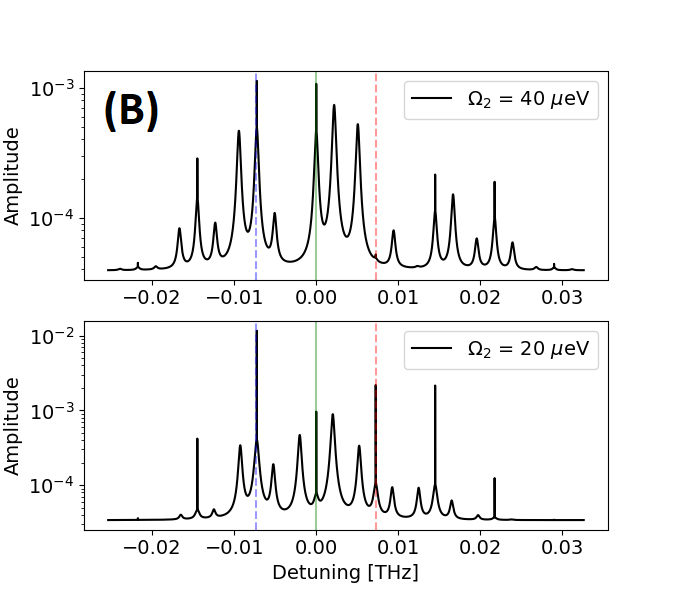}}
\end{subfigure}
       \caption{\textbf{a} A figure from \cite{gustinHighresolutionSpectroscopyQuantum2021} showing the emission two singular emission spectra for the bichromatic 2LS. The top spectrum uses $\Omega_2 = 40 \; \mu$eV, while the bottom is found using $\Omega_2 = 20 \; \mu$eV. The Orange line is the spectrum found using Gustin's bespoke simulation, and the black line is experimental data found by Gustin. \textbf{b} The FLiMESolve simulated results of the bichromatic system using (\textbf{top}) $\Omega_2 = 40 \; \mu$eV and (\textbf{bottom}) $\Omega_2 = 20 \; \mu$eV.}
        \label{FIG: Gustin_Spectra}
\end{figure*}

In all simulations, $\Omega_1$ is set to be $\frac{\Omega_1}{2\pi} = 30 \mu$eV Fig.~\ref{FIG: Gustin_Spectra} \textbf{a} and \textbf{b} show singular emission spectra for two separate values of $\Omega_2$. The top spectrum in each plot is found from $\frac{\Omega_2}{2\pi} =  40 \; \mu$eV, and the bottom plot in both is found using $\frac{\Omega_2}{2\pi} = 20 \; \mu$eV. Additionally, in the FLiMESolve-produced plots, three vertical lines are added to show the locations of the Mollow triplet sidebands that would be produced by $\Omega_1$ alone \cite{mollowPowerSpectrumLight1969a}.

On the  $\Omega_2 = 20 \; \mu$eV plots, the "original" Mollow triplet bands produced by Laser 1 are still largely visible. However, the main peak of the Mollow triplet has diminished, and the right sideband peak is also small compared to some new peaks around the original Mollow triplet peak locations. These new peaks are created by the second laser, which is detuned exactly to the left Mollow triplet peak. This second laser causes each original Mollow triplet peak to split into its own triplet. This can be seen by the peaks to the left and the right of the main Mollow triplet peaks which are highlighted by the vertical lines in the FLiMESolve plots. Overall, the peak locations and relative sizes match up very well between FliMESolve and Gustin's bespoke simulations for this specific setup.

Moving on to the upper half (or the $\Omega_2 =  40 \; \mu$eV parts of each image), the "new" triplet peaks begin to overtake the main Mollow triplet peaks. What at first appears to be a shifted triplet in Gustin's figure, located at around $10 \; \mu$eV on the x axis, reveals itself to actually be formed of the new sidebands from the "main" Mollow triplet as viewed in the FLiMESolve plots. The relative peak sizes and the exact peak frequencies match up with Gustin's results almost exactly, barring the peaks with the same spectral width as the laser. These peaks are likely laser scattering lines from the simulated quantum dot.

Finally, the speed at which MESolve and FLiMESolve are able to simulate the system is once again compared.
\begin{center}
\begin{table}[t]
\resizebox{\columnwidth}{!}{
\begin{tabular}{||c c c c c||} 
    \hline
    \multicolumn{1}{|p{2cm}|}{\centering Number of Periods} & \multicolumn{1}{|p{2cm}|}{\centering FLiMESolve Solution Time (s)} & \multicolumn{1}{|p{2cm}|}{\centering MESolve  Solution Time (s)} & \multicolumn{1}{|p{2cm}|}{\centering Solution Time Quotient} & \multicolumn{1}{|p{2cm}|}{\centering  Total Run Time Quotient}  \\
    \hline\hline
    $10^{1}$ & $1.2\pm0.6\times10^{-2}$ & $6\pm7\times10^{-3}$ & 2.41  & 0.49 \\ 
    \hline
    $10^{2}$ & $7\pm11\times10^{-2}$ &  $5\pm2\times10^{-2}$ & 2.26  & 0.70 \\ 
    \hline
    $10^{3}$ & $1.5\pm0.08\times10^{-1}$ & $0.49\pm0.02\times10^{-1}$ & 3.24 & 0.78  \\ 
    \hline
    $10^{4}$ & $1.67\pm0.02$ & $5.04\pm0.19$ & 3.01 & 0.79 \\
    \hline
    $10^{5}$ & $16.47\pm0.49$ & $49.33\pm1.33$ &  2.99 & 0.79\\
    \hline
\end{tabular}
}
 \caption{Solution time and total run time in seconds for FLiMESolve and MESolve for the bichromatic two level system.}
    \label{TABLE: Table2}
\end{table}
\end{center}

\subsection{$\pi$-Pulsed Two-Level System}

\begin{figure}[!b]
  \begin{minipage}{0.48\textwidth}
    \centering
    {\includegraphics[width=1\textwidth]{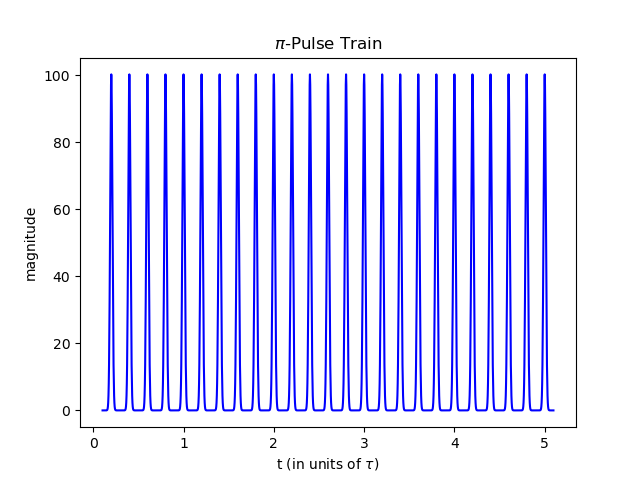}}
  \end{minipage}\hfill
        \caption{The pulse train used to drive the 2LS. The magnitude of each pulse is meaningless besides the fact that each pulse is area-normalized to $\pi$. The pulse train is set up to pulse five times per lifetime of the system.}
        \label{FIG: PulseTrain }
\end{figure}

The final system of interest is again the 2-level system, however, this time, the excitation is a train of $\pi$-pulses. The effect of the pulses is to invert the populations of the ground and excited states, with the system evolving as a non-driven 2LS between pulses. To begin, consider the rotating wave approximation form of the Hamiltonian given by
\begin{equation} \label{EQ:Pulsed_2LS_Hamiltonian}
\hat{H}(t) =\frac{1}{2}\begin{bmatrix}
-\Delta & 0 \\
0       & \Delta 
\end{bmatrix}
+
\frac{1}{2}\begin{bmatrix}
0 & 1 \\
1 & 0 
\end{bmatrix}\sum_{n=0}^{\infty}\frac{1}{\sigma \sqrt{2\pi}}e^{-\frac{(t-nT)^2}{2\sigma^2}}
\end{equation}
In the above equation, the period $T$ and the Gaussian standard deviation $\sigma$ are quantified relative to the lifetime of the system, defined by $T_1 = 2$ units of time, where the exact units don't matter because the Hamiltonian is set up to be scale invariant. The period is chosen to be $T = T_1/20$, and the Gaussian width is chosen to be $\sigma = T_1/16$ This has the result of creating five pulses within any given lifetime of the system. The resulting pulse train can be seen in Fig.~\ref{FIG: PulseTrain }.
\subsubsection{Relaxed Secular Approximation}
\begin{figure}[!t]
   \begin{minipage}{0.48\textwidth}
    \centering
    {\includegraphics[width=1\textwidth]{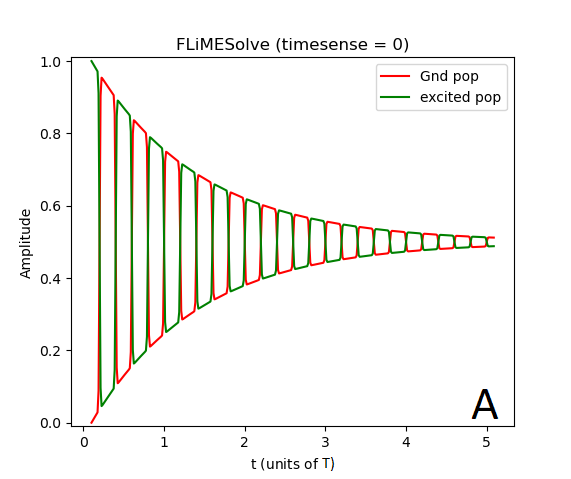}}
  \end{minipage}\hfill
  \begin{minipage}{0.48\textwidth}
    \centering
    {\includegraphics[width=1\textwidth]{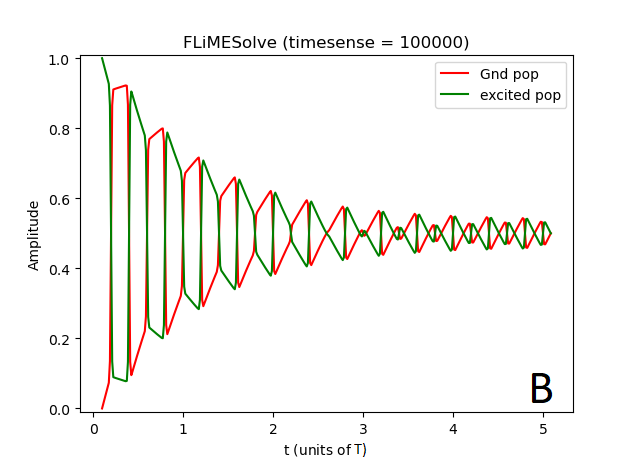}}
  \end{minipage}\hfill
        \caption{The state evolution as solved by FLiMESolve for (a) the most restrictive secular approximation and (b) effectively no secular approximation}
        \label{FIG:PulsedStateEvols}
\end{figure}

The solved behavior of this system looks markedly different depending on the level of secular approximation used. In Fig.~\ref{FIG:PulsedStateEvols}(a), the state evolution is plotted with the most restrictive secular approximation. 
In this case, it appears that the overall ``envelope'' of the decay is correct, however some nonphysical state evolution is also present. Rather than the excited state always decaying and the ground state always growing in population between the excitation pulses, the state with higher population is always decaying, and the state that has less population is always growing. This is clearly incorrect, as the system should act as a free 2LS between pulses, i.e., the excited state should always decay between pulses and the ground state should always grow between pulses.
However, in Fig.~\ref{FIG:PulsedStateEvols}(b), a more physically correct behavior can be seen. This figure shows the result of FLiMESolve handling the state evolution with an effectively infinite cutoff for the secular approximation. This figure shows much more physically accurate evolution, with the expected decay (growth) of the excited (ground) state population between pulses. 

Finally and once more, the solution times for this system are compared between MESolve and FLiMESolve. Contrary to the earlier examples, FLiMESolve is faster than MESolve for low numbers of simulation periods, but becomes slower as the number of simulated periods is increased. This is likely due to the fact that the secular approximation must be relaxed to achieve accurate results in the case of FLiMESolve. Because the resulting dynamics are dependent on many more frequencies than just the central frequency of the system, FliMESolve's strengths can't be leveraged as well and the solution method takes longer. However, the ability of FLiMESolve to relax to allow for more complicated time dependence is a noteworthy benefit to the solution method.

\begin{center}
\begin{table}[t]
\resizebox{\columnwidth}{!}{
\begin{tabular}{||c c c c c||} 
    \hline
    \multicolumn{1}{|p{2cm}|}{\centering Number of Periods} & \multicolumn{1}{|p{2cm}|}{\centering FLiMESolve Solution Time (s)} & \multicolumn{1}{|p{2cm}|}{\centering MESolve  Solution Time (s)} & \multicolumn{1}{|p{2cm}|}{\centering Solution Time Quotient} & \multicolumn{1}{|p{2cm}|}{\centering  Total Run Time Quotient}  \\
    \hline\hline
    $10^{1}$ & $0.021\pm.0.01$ & $0.15\pm0.014$ & 7.29  & 5.64 \\ 
    \hline
    $10^{2}$ & $0.12\pm0.007$ &  $0.49\pm0.015$ & 3.79  & 2.77 \\ 
    \hline
    $10^{3}$ & $0.97\pm0.04$ & $0.71\pm0.04$ & 0.73 & 0.047  \\ 
    \hline
    $10^{4}$ & $9.02\pm0.1$ & $2.63\pm0.07$ & 0.29 & 0.19 \\
    \hline
    $10^{5}$ & $89.12\pm0.16$ & $21.27\pm0.69$ &  0.23 & 0.16 \\
    \hline
\end{tabular}
}
 \caption{Solution time and total run time in seconds for FLiMESolve and MESolve for the $\pi$-pulsed two level system.}
    \label{TABLE: Table2}
\end{table}
\end{center}

%(Add something here comparing the performance of FLiMESolve and MESolve handling this system. Which one is faster? %By how much? You can also use a plot of MESolve results to confirm that FLiMESolve produces reliable results %without the secular approximation.)

\section{Discussion}
In this paper, a new type of ODE solver based on Floquet theory is discussed. A short derivation was presented, in which both a secular and nonsecular master equation of the Floquet-Lindblad type are derived. The secular version is easier to obtain and cheaper to solve computationally, but there are cases where a relaxed secular approximation is necessary.
After discussing the theory behind the FLiME, two exemplar systems were discussed. While these systems are themselves rather simple, they show the utility of this new solver. The first system was that of a driven 2-level system, however, instead of making the usual rotating wave approximation, the full Hamiltonian was used due to high driving power. FLiMESolve was able to address the computational overhead required of this time-dependent Hamiltonian via its efficiency and its applicability to problems in the Floquet regime of ordinary differential equations. The second system was that of the 2-level system excited by a pulse train. Since the pulses are made up of a sum of waves of different harmonic frequencies, the full secular approximation realizes a state evolution that is nonphysical because it ignores frequencies above the fundamental. However, the ability of FLiMESolve to relax the secular approximation allows for the recovery of accurate system evolution.

There are two avenues for future work on FLiMESolve. Firstly, the decay rate can be made non-constant over all frequencies. This would bring the form of the decay rate, $\gamma$, to match non-secular master equations derived elsewhere, and would allow for the simulation of even more richly dynamic systems \cite{liaoLindbladRedfieldForms2021}. The second avenue for future work is to incorporate multimode Floquet theory into the FLiME \cite{poertnerValidityManymodeFloquet2020,hoSemiclassicalManymodeFloquet1983a}. In semiclassical multimode Floquet theory, commensurate driving frequencies (i.e,. those whose quotient is a rational number) are themselves treated as the Fourier expansion of some lower frequency driving term, given by the lowest common denominator of the two (or more) commensurate frequencies. This further complicates the indexing of any eventual master equations, but would allow for the simulation of systems with many commensurate driving terms.

\bibliography{References}

\end{document}